\documentstyle[prb,aps]{revtex}

\begin{document}

\draft  \tighten

\preprint{MA/UC3M/18/94}

\title{Thomas-Fermi approach to resonant tunneling in $\delta$-doped
diodes}

\author{Enrique Diez,$^{\dag}$ Francisco Dom\'{\i}nguez-Adame,$^{\ddag}$
and Angel S\'{a}nchez$^{\dag}$}

\address{$^{\dag}$Escuela Polit\'{e}cnica Superior, Universidad
Carlos III de Madrid, c./ Butarque 15 \\ E-28911 Legan\'{e}s,
Madrid, Spain\\
$^{\ddag}$Departamento de F\'{\i}sica de Materiales,
Facultad de F\'{\i}sicas, Universidad Complutense,\\
E-28040 Madrid, Spain}

\maketitle

\begin{abstract}

We study resonant tunneling in B-$\delta$-doped diodes grown by
Si-molecular beam epitaxy.  A Thomas-Fermi approach is used to obtain
the conduction-band modulation.  Using a scalar Hamiltonian within the
effective-mass approximation we demonstrate that the occurrence of
negative differential resistance (NDR) only involves conduction-band
states, whereas interband tunneling effects seem to be negligible.  Our
theoretical results are in very good agreement with recent experimental
observations of NDR in this type of diodes.

\end{abstract}



\narrowtext

\section{Introduction}

Resonant tunneling (RT) through semiconductor diodes has recently
attracted considerable attention because of its applications in
ultra-high-speed electronic devices.  Their most remarkable feature
is that their $I$-$V$ characteristics show
negative differential resistance (NDR).  Most of these devices are based
in double-barrier structures, where RT occurs via quasi-bound states
within the well region in the same energy band (conduction- or
valence-band).  On the contrary, other kind of RT devices present
interband transitions as, for instance, InAs-GaSb or InAs-AlSb-GaSb
structures. \cite{Wang} In the later case, Kane's parameter is not
negligible as compared with bandgaps and, consequently, more elaborated
band-structures are required to fully account for RT.

Recently, Sardela
{\em et al.} \cite{Sardela}
have observed NDR at room temperature in B-$\delta$-doped
diodes grown by Si-molecular beam epitaxy. A schematic
cross section of the studied device in shown in Fig.~\ref{fig1}.
Experimental NDR peaks are seen to
appear at about $\pm 0.2\,$V.  However, it is not
clear whether RT occurs via intraband or interband mechanisms: Indeed,
the
authors argued that preliminary calculations based on the static
solution of the Poisson's equation show that the whole structure only
supports a very shallow quasi-bound state and, consequently,
conduction-band modulation cannot explains the observed NDR. Hence, they
claimed that the valence-band plays a major role in RT of these diodes
and therefore two-band Hamiltonians \cite{Bastard} are necessary to
understand the obtained results.

In this paper we will show that scalar Hamiltonians for the
conduction-band envelope functions work well to explain the results of
Sardela {\em et al.} \cite{Sardela}  Instead of using self-consistent
approximations to compute the one-electron potential due to
$\delta$-doping, we will concern ourselves with the nonlinear
Thomas-Fermi (TF) formulation as introduced by Ioratti.\cite{Ioratti}
This approach reproduces accurately the various subbands densities in
periodically $\delta$-doped GaAs layers at moderate and high doping
\cite{Egues} and it has proved itself useful n describing the
electronic structure in the presence of applied external fields.
\cite{SST,PRB} We thus obtain
results showing the existence of several quasi-bound
states with large lifetimes localized between the two $\delta$-layers
that successfully explain the $I-V$ characteristics previously
obtained by Sardela {\em et al.}

\section{Model}

We consider an electron of effective mass $m^*$ in the $\delta$-doped
diode (see Fig.~\ref{fig1}) in the presence of an electric field applied
along the growth direction.  Nonparabolicity effects are neglected so
that $m^*$ will be taken to be independent of the electron energy.  The
electron wave function and the total energy of the electron are given by
$\Psi({\bf r},{\bf k}_\perp)=\exp(i{\bf k}_\perp\cdot{\bf r}_\perp)
\phi(x)$ and $E({\bf k}_\perp)=E+(\hbar^2{\bf k}_\perp^2/2m^*)$,
where $\phi(x)$ is the envelope function.  Note that we have taken
isotropic bands at the $X$ minima, and, in fact, the band structure of Si is
strongly anisotropic ($m^*_l/m^*_t \sim 5$).  Hence, we should consider
an {\em average} value $m^*$ that would be obtained, for example, in a
measurement of mobility.\cite{Jaros} In the effective-mass approximation
the envelope function satisfies the following Schr\"odinger equation
\begin{equation}
\left[-\,{\hbar^2\over 2m^*}\,{d^2\phantom{x}\over dx^2}+
V_{TF}(x)-eFx\right]\,\phi(x)=E\,\phi(x).
\label{eq1}
\end{equation}
Here $F$ is the applied electric field and $V_{TF}(z)$ is the solution
of the nonlinear TF equation (see Refs.~\onlinecite{SST} and
\onlinecite{PRB} for details).  As usual in scattering problems, the
envelope function at the bottom electrode is a superposition of incident
and reflected traveling waves, whereas at the top one there is only a
transmitted wave.  Hence standard numerical techniques \cite{SST} can be
used to obtain the transmission coefficient $\tau(E,V)$ for a given
incident energy $E$ and a given applied voltage $V=FL$, $L$ being the
length of the whole structure.  Due to the high doping of the
electrodes, the electric field is assumed to be nonzero only within the
structure.  In addition, these high screening effects imply that
$V_{TF}(z)$ also vanishes at the electrodes.

\section{Results}

We have set an effective mass $m^*=0.33$ and a dielectric constant
$\epsilon=11.7$ in Si.\cite {Jaros} In the absence of external fields,
the TF potential presents two peaks corresponding to the
$\delta$-layers, as shown in Fig.~\ref{fig2}.  It is worth mentioning
that the maximum value of the conduction-band modulation is about $20\%$
of the bandgap in Si. This value is of the same order of that obtained
self-consistently in Sb-$\delta$-doped grown by Si-molecular beam epitaxy
with similar areal concentration of impurities.\cite{Radamson} Assuming
a similar profile for the valence-band, we are led to the conclusion
that interband RT should not play a significant role, at least at
moderate electric fields.  Hence, the scalar Hamiltonian (\ref{eq1})
suffices to describe the electronic structure of the diode.  In
addition, since we are close to the conduction-band edge, parabolic
subbands hold, as we assumed previously.  To elucidate whether the
conduction-band modulation can support narrow quasi-bound states, we
have numerically evaluated the transmission coefficient at zero bias
$\tau(E,0)$ as a function of the incident energy and results are plotted
in Fig.~\ref{fig3}.  From this figure it becomes clear that four
resonances (quasi-bound states) appear below the top of the potential.
The levels are nearly equally spaced in this case: This is easy to
understand if we notice that the potential profile between the two
$\delta$-layers is almost parabolic.  Interestingly, we thus have
demonstrated that $\delta$-doped diodes could be used to achieve equally
spaced peaks in the collector characteristics, in analogous way to
parabolic wells formerly proposed by Capasso and Kiehl.\cite{Capasso}

The tunneling current density at a given temperature $T$ for the diode
sketched in Fig.~\ref{fig1} can be calculated within the
stationary-state model from
\begin{mathletters}
\label{eq2}
\begin{equation}
j(V)={m^*ek_BT\over 2\pi^2\hbar^3}\,\int_0^\infty\> \tau(E,V)N(E,V)\,dE,
\label{eq2a}
\end{equation}
where $N(E,V)$ accounts for the occupation of states to both sides of
the device, according to the Fermi distribution function, and it is
given by
\begin{equation}
N(E,V)=\ln\left(\frac{1+\exp[(E_F-E)/k_BT]}{1+\exp[(E_F-E-eV)/k_BT]}
\right)
\label{eq2b}
\end{equation}
\end{mathletters}

We show in Fig.~\ref{fig4} the computed $j-V$ characteristics at
$T=77\,$K and at room temperature.  The curves have been obtained taking
the Fermi energy at the conduction-band edge away from the
$\delta$-layers.  Other values of the Fermi energy simply modify the
scale of $j$, keeping its shape unchanged.  The main NDR feature,
also observable at room temperature, appears at about $0.22\,$V,
which is very close to experimental results of Sardela {\em et al.}
\cite{Sardela}
Note that our computation predicts two separated peaks in the $j-V$
characteristics at around $0.22\,$V, while experiments show only a single,
broader peak.  Inelastic scattering mechanisms, not included in our
analysis, are known to
cause a broadening of the intrinsic level width, and the
amount of inelastic \cite{Booker} broadening makes the two
separated resonances to
merge into a single one, according to experimental results.  Moreover,
there exist smaller peaks at about $0.10\,$V and $0.15\,$V, arising from
the lower resonances of the well.  In experiments \cite{Sardela} such
small peaks are almost unnoticeable, probably because we are
overestimating its height due to the absence of inelastic effects in our
model which, of course, tend to disrupt coherent tunneling.  The
theoretical peak-to-valley ratio of the main NDR feature is $2.8$ at
$77\,$, larger than the experimental value $1.4$ at the same
temperature.  The reason why the
theoretical calculations predicts larger
peak-to-valley ratios is that sequential tunneling is not being
considered.

\section{Summary}

We have proposed a simple model to explain RT through
$\delta$-doped diodes grown by Si-molecular beam epitaxy.  The
conduction-band modulation is obtained by means of the TF approach and
the corresponding electronic states are described by a scalar
Hamiltonian.  Our results show that NDR effects are due to
conventional resonant tunneling process in vertical transport, whereas
interband tunneling does not give rise to significant contributions.
The obtained results are in excellent agreement
with recent measurements by Sardela {\em et
al.}\cite{Sardela} This success suggests that our approach, being very
simple and computationally inexpensive, may be very useful in dealing
with semiconductor nanostructures.

\acknowledgments

F.\ D-A.\ acknowledges support from UCM through project PR161/93-4811.
A.\ S.\ acknowledges partial support from C.I.C.\ y T.\ (Spain) through
project PB92-0248 and by the European Union Human Capital and Mobility
Programme through contract ERBCHRXCT930413.

\begin{figure}
\caption{Schematic cross section of a B-$\delta$-diode grown by
Si-molecular beam epitaxy, as studied in
Ref.~\protect{\onlinecite{Sardela}}. The whole structure is embebded
between two electrodes consisting of highly $n$-doped Si layers. Each
$\delta$-layer is $8\,$\AA\ width with an acceptor concentration
$1.4\times 10^{13}\,$cm$^{-2}$.}
\label{fig1}
\end{figure}

\begin{figure}
\caption{Modulation of the conduction-band obtained from the
Thomas-Fermi approach for the structure shown in
Fig.~\protect{\ref{fig1}} without applied electric field.}
\label{fig2}
\end{figure}

\begin{figure}
\caption{Transmission coefficient without applied electric field as a
function of the incident energy.}
\label{fig3}
\end{figure}

\begin{figure}
\caption{Computed $j-V$ characteristics of $\delta$-doped diodes at (a)
$T=77\,$K and (b) room temperature.}
\label{fig4}
\end{figure}


\end{document}